\documentclass[twocolumn]{aastex62}

\NoNewPageAfterKeywords

\usepackage[utf8]{inputenc}
\DeclareUnicodeCharacter{2212}{-}
\usepackage{graphicx}
\graphicspath{{./figures/}{./}}
\usepackage{amsmath}
\usepackage{afterpage}
\usepackage{enumitem}
\setenumerate{itemsep=0mm}
\hypersetup{
  urlcolor        = {blue},
}
\usepackage{comment}
\usepackage{multirow}

\newcommand{\northernname}{\textsl{North-East~Arc}\xspace}

\newcommand{\typeab} {\mbox{\emph{ab}}\xspace}

\newcommand{\documentname}{\textit{Letter}}

\usepackage{lineno}

\usepackage{soul}
\usepackage{amsmath}
\usepackage{amssymb}
\usepackage{xspace}
\usepackage{xifthen}
\usepackage{eso-pic}

\definecolor{forestgreen}{HTML}{228B22}
\definecolor{urlblue}{HTML}{000000}

\newcommand{\response}[1]{{}}

\newcommand{\Gaia}{{\it Gaia}\xspace}
\newcommand{\SSSSS}{${S}^5$\xspace}

\mathchardef\mhyphen="2D

\newcommand{\roughly}{\ensuremath{ {\sim}\,} }

\newlength{\dhatheight}

\newcommand{\code}[1]{\texttt{#1}\xspace}

\newcommand{\unit}[1]{\ensuremath{\mathrm{\,#1}}\xspace}

\newcommand{\Gyr}{\unit{Gyr}}
\newcommand{\Myr}{\unit{Myr}}

\newcommand{\degree}{\ensuremath{{}^{\circ}}\xspace}

\newcommand{\masyr}{\unit{mas}~\unit{yr}^{-1}\xspace}

\newcommand{\km}{\unit{km}}
\newcommand{\kms}{\km \second^{-1}}

\newcommand{\kpc}{\unit{kpc}}
\newcommand{\second}{\unit{s}}

\newcommand{\tabref}[1]{Table~\ref{tab:#1}}

\newcommand{\figref}[1]{Figure~\ref{fig:#1}}

\newcommand{\bandvar}[2][]{%
  \ifthenelse{\isempty{#1}}{\var{#2}}{\var{#2\_#1}}%
}

\newcommand{\modulus}{\ensuremath{m - M}\xspace}

\newcommand{\ra}{{\ensuremath{\alpha_{2000}}}\xspace}
\newcommand{\dec}{{\ensuremath{\delta_{2000}}}\xspace}
\newcommand{\age}{{\ensuremath{\tau}}\xspace}

\newcommand{\feh}{{\ensuremath{\rm [Fe/H]}}\xspace}

\newcommand{\ugali}{\code{ugali}}
\newcommand{\var}[1]{\ensuremath{\texttt{\MakeUppercase{#1}}}\xspace}

\providecommand\physrep{\ref@jnl{Phys.~Rep.}}%
\providecommand\apjs{\ref@jnl{ApJS}}%
\providecommand{\jcap}{\ref@jnl{JCAP}}%

\begin{document}

\title{Discovery of Extended Tidal Tails around the Globular Cluster Palomar 13}

\author[0000-0003-2497-091X]{Nora~Shipp}
\affiliation{Department of Astronomy and Astrophysics, University of Chicago, Chicago IL 60637, USA}
\affiliation{Kavli Institute for Cosmological Physics, University of Chicago, Chicago, IL 60637, USA}
\affiliation{Fermi National Accelerator Laboratory, PO Box 500, Batavia, IL 60510, USA}

\author[0000-0003-0872-7098]{Adrian~M.~Price-Whelan}
\affiliation{Center for Computational Astrophysics, Flatiron Institute, Simons Foundation, 162 Fifth Avenue, New York, NY 10010, USA}

\author[0000-0001-6584-6144]{Kiyan Tavangar}
\affiliation{Department of Astronomy and Astrophysics, University of Chicago, Chicago IL 60637, USA}

\author[0000-0002-6330-2394]{Cecilia~Mateu}
\affiliation{Departamento de Astronom\'ia, Facultad de Ciencias, Universidad de la Rep\'ublica, Igu\'a 4225, 14000, Montevideo, Uruguay}

\author[0000-0001-8251-933X]{Alex~Drlica-Wagner}
\affiliation{Fermi National Accelerator Laboratory, PO Box 500, Batavia, IL 60510, USA}
\affiliation{Kavli Institute for Cosmological Physics, University of Chicago, Chicago, IL 60637, USA}
\affiliation{Department of Astronomy and Astrophysics, University of Chicago, Chicago IL 60637, USA}

\email{norashipp@uchicago.edu}

\begin{abstract}
We use photometry from the DECam Legacy Survey to detect candidate tidal tails extending $\roughly 5 \deg$ on either side of the Palomar 13 globular cluster.
The tails are aligned with the proper motion of Palomar 13 and are consistent with its old, metal-poor stellar population.
We identify three RR Lyrae stars that are plausibly associated with the tails, in addition to four previously known in the cluster.
From these RR Lyrae stars, we find that the mean distance to the cluster and tails is $23.6 \pm 0.2 \kpc$ and estimate the total (initial) luminosity of the cluster to be $L_V=5.1^{+9.7}_{-3.4}\times 10^3 L_\odot$, consistent with previous claims that its initial luminosity was higher than its current luminosity.
Combined with previously-determined proper motion and radial velocity measurements of the cluster, we find that Palomar 13 is on a highly eccentric orbit ($e\sim 0.8$) with a pericenter of $\roughly 9\kpc$ and an apocenter of $\roughly 69\kpc$, and a recent pericentric passage of the cluster $\roughly 75 \Myr$ ago.
We note a prominent linear structure in the interstellar dust map that runs parallel to the candidate tidal features, but conclude that reddening due to dust is unlikely to account for the structure that we observe.
If confirmed, the Palomar 13 stellar stream would be one of very few streams with a known progenitor system, making it uniquely powerful for studying the disruption of globular clusters, the formation of the stellar halo, and the distribution of matter within our Galaxy.
\end{abstract}


\section{Introduction}
\label{sec:intro}

Stellar streams---remnants of tidally-disrupted star clusters and satellite galaxies---provide information about the formation history and dark matter distribution of the Milky Way \citep[e.g.,][]{Johnston:1998, Helmi:1999, Bonaca:2018}.
Dynamically cold streams from disrupting star clusters are also extremely sensitive to gravitational perturbations from massive substructures \citep[e.g.,][]{Erkal:2015b, Bonaca:2019b}.
The study of stellar streams is thus a promising avenue for studying the distribution of dark matter at sub-galactic scales.

From our perspective in the Galaxy, the discovery and study of stellar streams requires wide-field sky surveys \citep[e.g.,][]{Rockosi:2002, Majewski:2003}.
The population of known streams has increased substantially due to deep, well-calibrated photometric surveys such as the Dark Energy Survey (DES; \citealt{DES:2016}; see \citealt{Shipp:2018}), and our ability to characterize stream motions has been recently revolutionized by the \Gaia mission (\citealt{Gaia:2018}; see, e.g., \citealt{Price-Whelan:2018, Malhan:2018b, Shipp:2019}) and coordinated spectroscopic follow-up \citep[e.g., the \SSSSS survey;][]{Li:2019}.

The known population of stellar streams now comprises  $\roughly70$ candidates distributed throughout the Milky Way stellar halo.\footnote{\url{https://github.com/cmateu/galstreams}}
Most stellar streams likely originated from globular clusters, but the vast majority have no known surviving progenitors.
In contrast, many globular clusters have extended extra-tidal envelopes \citep[e.g.,][]{Kuzma:2018}, but do not appear to host dense tidal tails.
It is currently unclear how this discrepancy connects to the surface brightness evolution of tidal tails \citep[e.g.,][]{Balbinot:2018}.

A stark exception to these statements is the globular cluster Palomar 5 (Pal 5) and its tidal tails \citep[e.g.,][]{Rockosi:2002}, which span $\roughly 25\degr$ on the sky \citep[e.g.,][]{Bonaca:2020} and contain a total stellar mass comparable to the surviving cluster \citep[e.g.,][]{Ibata:2017}.
The Pal 5 system is an archetype of stellar stream formation and has provided dynamical constraints on the dark matter distribution around the Milky Way \citep[e.g.,][]{Kuepper:2015}.
However, recent studies have found that the Pal~5 orbit and stream may be perturbed by the time-dependent influence of the Galactic bar \citep{Pearson:2017} and/or massive substructures \citep{Erkal:2017}.
These realizations motivate the study of other globular cluster streams as a way to disentangle the complex dynamical phenomena affecting streams in the inner Galactic halo.

In order to increase the sample of stellar streams with identified progenitors, we have initiated a systematic search for tidal debris structures using photometric data from the DESI Legacy Imaging Surveys \citep{Dey:2019}.
A first result from this search is the detection of extended tidal structures associated with the globular cluster Palomar 13 (Pal 13).
Pal 13 is a low-luminosity globular cluster  \citep[$M_V \sim -2.8$;][]{Bradford:2011} at a Galactocentric distance of $\roughly 25\kpc$ \citep{Cote:2002}.
Several studies have previously suggested that Pal 13 is undergoing tidal disruption because of the spatial distribution of its blue straggler population \citep{Siegel:2001}, its large radial velocity dispersion \citep{Cote:2002}, and its extended radial profile \citep[$r_{1/2} = 1.27' \pm 0.16'$;][]{Bradford:2011}.
More recently, it was shown that Pal~13 member stars display a significant proper motion scatter \citep[using \Gaia DR2 astrometry;][]{Yepez:2019}, and its stellar population extends to almost twice the estimated Jacobi radius of $\sim$5–-10' (using photometry from the Dark Energy Camera Legacy Survey, DECaLS; \citealt{Piatti:2020}).
In this \documentname, we analyze DECaLS data over a large region around Pal 13 and present evidence for a linear debris structure aligned with the proper motion and orbit of the cluster, extending $\roughly 5\degr$ in either direction from the cluster center.

\begin{figure*}[th!]
    \centering
    \includegraphics[width=\textwidth]{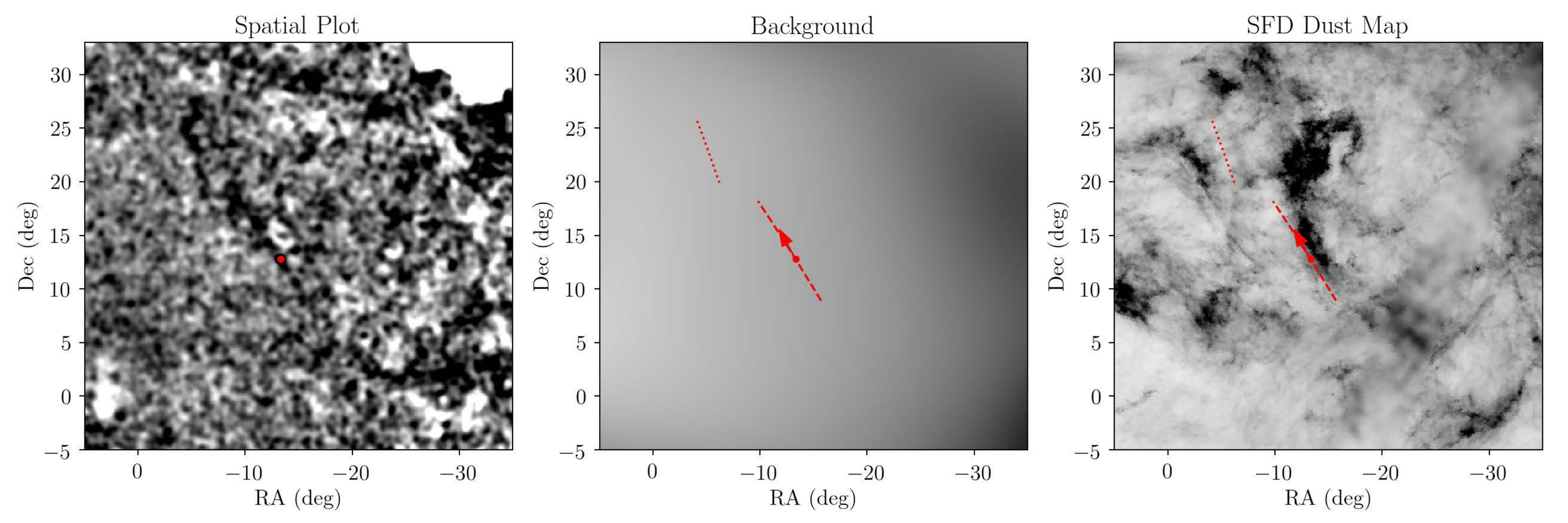}

\caption{\textit{Left}: Spatial map of the isochrone-selected residual stellar density at the distance modulus of Pal 13 ($m-M=16.8$). \textit{Middle}: Smooth polynomial background model in the area around Pal~13. \textit{Right}: The SFD dust map in the area around Pal~13.
The red dot represents the location of the Pal~13 cluster, the red dashed line traces the path of the great circle between the endpoints listed in \tabref{params}, and the red dotted line traces the path of a great circle along the \northernname.
The red arrow indicates the proper motion of Pal~13 from \citet{Vasiliev:2019}. An animated version of the left-hand panel can be found \href{https://home.fnal.gov/~kadrlica/movies/pal13paper_v1.gif}{at this url}; a wider-area animation can be found \href{https://home.fnal.gov/~kadrlica/movies/pal13widepaper_v1.gif}{at this url}.
}
\label{fig:spatial}
\end{figure*}

\section{Data \& Analysis}
\label{sec:data}

We perform our search using DECaLS data distributed in Data Release 8 (DR8) of the DESI Legacy Imaging Surveys \citep{Dey:2019}.
Source detection is performed with a PSF- and SED-matched-filter detection on the stacked images, with a $6\sigma$ detection limit.
Morphological fitting and photometry are performed with the \code{Tractor} code initialized at the positions of detected sources \citep{Lang:2016}, and we consider sources with \texttt{TYPE = `PSF'} to be stars.
We perform our search on the $g, r, z$-band data, and we require that sources are detected in all three bands.
In addition, we require that sources pass quality cuts: \texttt{(ANYMASK\_G == 0) \& (ANYMASK\_R == 0) \& (ANYMASK\_Z == 0)} and \texttt{(FRACFLUX\_G < 0.05) \& (FRACFLUX\_R < 0.05) \& (FRACFLUX\_Z < 0.05)}.
We correct the measured fluxes for interstellar extinction using the provided  extinction values (which make use of \citealt{Schlafly:2011}).

We perform an unweighted matched-filter search to this data in color--magnitude space \citep[e.g.,][]{Rockosi:2002}, following the procedure outlined in Section 3.1 of \citet{Shipp:2018}.
We select stars around a synthetic isochrone from \citet{Dotter:2008} and implemented in \ugali \citep{Bechtol:2015, Drlica-Wagner:2015}, with an age of 13.5 Gyr and a metallicity of $Z = 0.0001$ ($\feh = -2.2$).
The selection is defined as in Equation (4) of \citet{Shipp:2018}, with the selection width parameters determined empirically by comparison to old, metal-poor stellar populations in the data, such as globular clusters and dwarf galaxies (our initial search is not specifically tailored to Pal~13).
We limit our isochrone selection to an absolute magnitude of $M_g > 3.5$ to select stars lying along the higher signal-to-noise main sequence.
We limit the observed magnitude range of our matched filter to $g < 23$ due to the decreased uniformity of the data at fainter magnitudes.
We perform this selection in both $g$ vs.\ $g - r$ and $g$ vs.\ $r - z$.

We scan our isochrone filter in distance modulus from $15 < m - M < 19$ ($10 < D < 63 \kpc$) in steps of 0.1 mag.
At each distance modulus step, we fit a 5th-order polynomial to the filtered stellar density as a function of $\ra,\dec$ and smooth the data by applying a Gaussian filter with a smoothing kernel of $0\fdg25$. We subtract the background polynomial model from this smoothed data and visually inspect the residual stellar density maps as shown in \figref{spatial}.
We find several known features in the stellar density, including the Triangulum stream \citep{Bonaca:2012} and the Sagittarius stream \citep{Majewski:2003}.
The most prominent unknown linear feature is coincident with the globular cluster Pal 13, centered at ($\ra,\dec = 346\fdg7, 12\fdg8$).

\begin{figure*}[t!]
    \centering
    \includegraphics[width=0.95\textwidth]{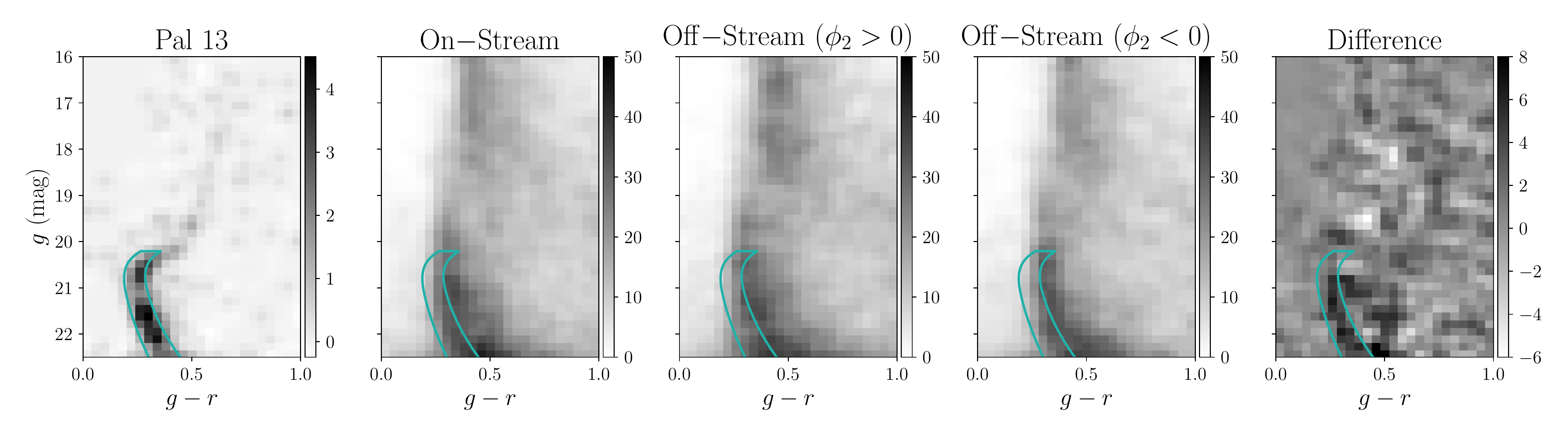}
    \caption{Color--magnitude diagrams for a region within $0\fdg4$ of the Pal 13 cluster (left), a region within 0\fdg25 of the track of the Pal~13 tidal features excluding the cluster (left middle), and a background region offset from the Pal~13 track by 2\fdg0 (right middle).
    The Hess difference diagram between the on-stream and off-stream regions is shown in the right-most panel.
    Over-plotted in cyan is the isochrone selection with $m - M = 16.8$, $\age = 13.5 \Gyr$, and $Z =0.0001$.}
    \label{fig:cmd}
\end{figure*}

\section{Extended Tidal Tails Around Pal 13}
\label{sec:tails}

In order to further examine the linear feature around Pal 13, we first determine optimized isochrone filter parameters (tuned to the stellar population of Pal~13) by-eye via comparison to the Pal~13 color--magnitude diagram in the left-hand panel of \figref{cmd}. We do not attempt to fit physically significant isochrone parameters to the cluster, and instead empirically determine a filter to most effectively select the Pal 13 signal (outlined in cyan in all panels of \figref{cmd}). We choose to broaden our filter in the blue direction in order to select stars at the blue edge of the main sequence, which typically provide the highest signal-to-noise in matched-filter searches.
We adopt isochrone parameters: $\modulus = 16.8$, \age = 13.5 Gyr, and $Z = 0.0001$.
The selection region is also defined by a magnitude broadening ($\Delta \mu = 0.5$), an asymmetric color broadening, $C_{1,2} = (-0.02, 1.0)$, and a multiplicative factor on the magnitude uncertainty ($E = 2$), which describes the error-dependent spread in color. All parameters are described in greater detail in Section 3.1 of \citet{Shipp:2018}.

The linear features detected by our isochrone filter extend $\sim 5 \degr$ in each direction from the center of Pal 13. We estimate the endpoints to be $(\ra, \dec) = (-9.8, 18.2)\degr$ and $(-15.7, 8.9)\degr$ and find that the track of the tails is well-matched by the great circle connecting these endpoints.
We define a new coordinate system such that the endpoints lie along the x-axis and the cluster is positioned at $(\phi_1, \phi_2) = (0, 0)\degr$.
The coordinate system is defined by the rotation matrix,
\begin{equation}
R=
\begin{bmatrix}
 0.94906836 & -0.22453560 & 0.22102719 \\
-0.06325861 &  0.55143610 & 0.83181523 \\
-0.30865450 & -0.80343138 & 0.50914675 \\
\end{bmatrix}
\quad ,
\end{equation}
which transforms from ICRS to stream-aligned coordinates.
Throughout the text, we consider the ``on-stream" region to be between the endpoints ($-4\fdg5 < \phi_1 < 6\fdg4$) and $-0\fdg25 < \phi_2 < 0\fdg25$, \emph{excluding} the region within 0\fdg4 of the globular cluster.

The region around Pal 13 is complicated by a linear feature in the interstellar dust adjacent to the globular cluster (right panel of \figref{spatial}).
Imperfect reddening corrections can lead to artificial color shifts that could increase or decrease the number of stars passing our isochrone filter.
Indeed, we see a deficit of stars to the northwest of Pal~13 coinciding with this dust feature.
A more subtle concern is that the underdensity of stars associated with the dust feature could conspire with the rapidly falling foreground stellar density to manifest as an apparent linear overdensity of stars directly adjacent to the dust lane.\footnote{This is the 2-dimensional analog of the well-known axiom in particle physics: when analyzing steeply falling spectra, ``every dip creates a bump''.}

\begin{figure}[t!]
    \centering
    \includegraphics[width=0.99\columnwidth]{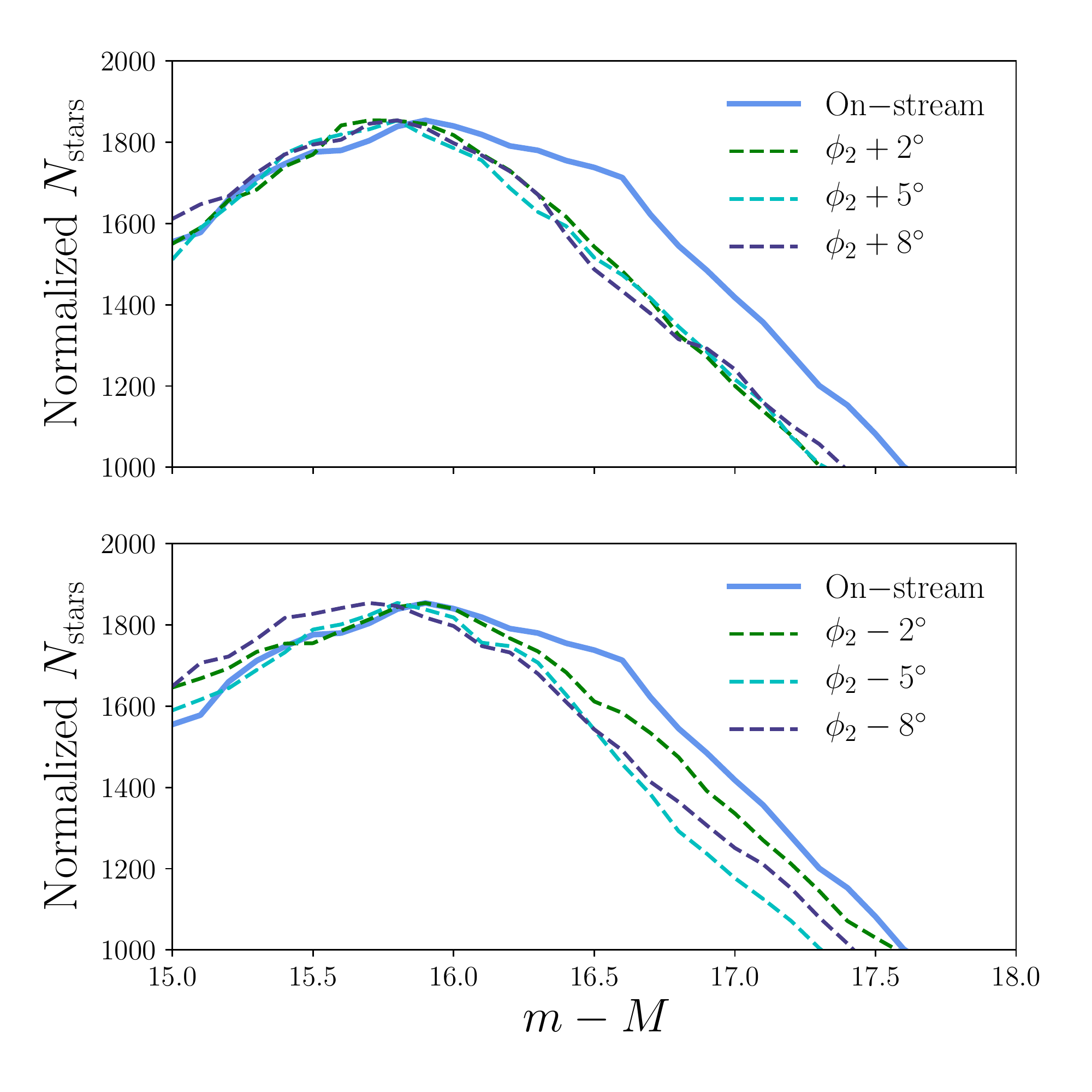}
    \caption{Relative number of stars as a function of distance modulus in the on-stream region (blue, solid line) and six equal-area regions offset by $\pm$ 2\degree, 5\degree, and 8\degr (dashed lines).
    The number of stars in the off-stream regions are normalized to the maximum value of the on-stream region to emphasize the difference in shape of the curves.
    The measured stellar density in the on-stream region decreases more slowly with increasing distance than the density in the off-stream regions, suggesting that the observed overdensity is distinct from the foreground population.}
    \label{fig:density}
\end{figure}

To distinguish a spurious density variation from the presence of a distant, metal-poor stellar population associated with Pal~13, we examine the significance of the tidal tails as a function of distance and compare to neighboring, off-stream regions.
If the tails are an artifact of the foreground population, they would have the same distance dependence as the neighboring regions, while tidal features associated with Pal~13 would be more prominent at larger distances.
In \figref{density}, we show the normalized number of stars along the tidal tails, and in three adjacent equal-area regions above and below the stream.
The number of stars falls off more slowly with distance modulus in the on-stream region than in any of the off-stream regions, including those overlapping with the dust feature visible in \figref{spatial}.
We conclude that the candidate tidal tails have a different distance dependence than the foreground stars and are more likely associated with the distant stellar population of Pal~13.

In addition, we note the presence of a second feature extending from $(\ra, \dec) = (-6\fdg2, 19\fdg9)$ to $(-4\fdg0, 26\fdg0)$, which we refer to as the ``\northernname'' (dotted red line in \figref{spatial}). The feature is disconnected from the observed Pal~13 tidal tails; however, the orientation of the two features bears some resemblance to known ``broken'' streams, such as the ATLAS and Aliqa Uma stream \citep{Li:2019}. Therefore, although we find no concrete evidence of an association between this \northernname and Pal 13, this feature may be worth investigating further.

\subsection{Color--Magnitude Diagram}

We examine the tidal features in color--magnitude space to characterize the stellar population and compare to the globular cluster itself.
\figref{cmd} shows binned color--magnitude diagrams in four different regions.
The first panel shows stars selected within 0\fdg4 of the Pal 13 globular cluster, the second panel shows stars within the on-stream region (excluding the cluster), and the third and fourth panels shows stars within equal-area off-stream regions, offset from the stream by $\pm 2\fdg0$ in $\phi_2$.
The fifth panel shows the background-subtracted Hess diagram of the on-stream region. The faint overdensity in the fifth panel is plausibly consistent with the globular cluster signal, and suggests that the cluster stellar population may in fact extend in the direction of the candidate tidal tails, which would support the association between the candidate tails and the Pal 13 globular cluster.
We also look for evidence of the \northernname in color--magnitude space and are unable recover a strong signal, making it difficult to claim association between the \northernname and Pal 13.

\newcommand{\paramscaption}{Measured properties of Pal~13 tidal tails.}
\begin{deluxetable}{l c}[h!]
\tablecolumns{13}
\tablewidth{0pt}
\tabletypesize{\scriptsize}
\tablecaption{ \paramscaption \label{tab:params}}
\tablehead{Parameter & Value}
\startdata
Endpoint (\ra, \dec) & $(-9.8, 18.2)\degr$ \\
Endpoint (\ra, \dec) & $(-15.7, 8.9)\degr$ \\
Length & $10.9\degr$  \\
Width & $0.25\degr$ \\
\enddata

\end{deluxetable}

\begin{figure*}[t!]
    \centering
    \includegraphics[width=0.85\textwidth]{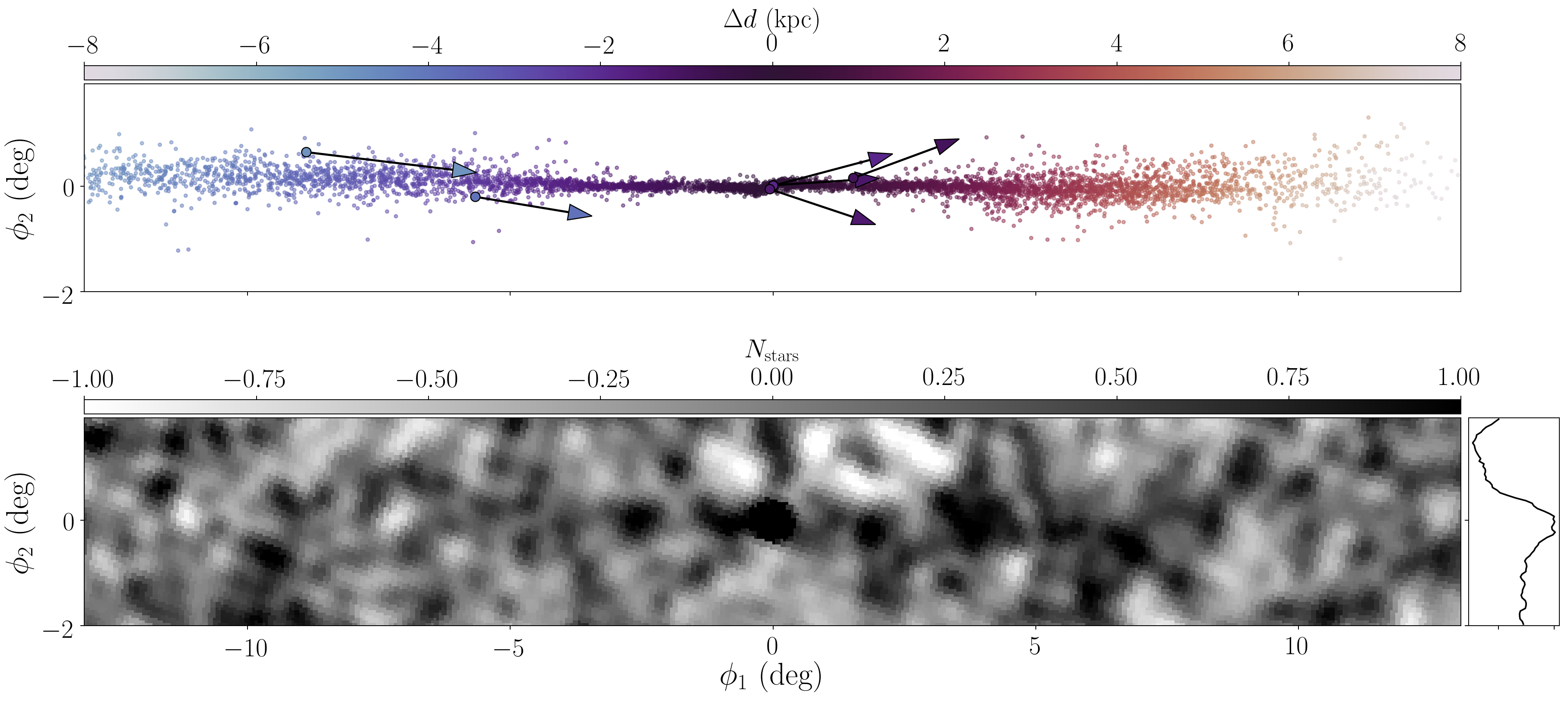}
    \caption{{\it Upper:} Star particles in a stream model generated from the best-fit orbit to the measured position and velocity of Pal~13, and the on-sky track of the tidal tails. Likely member RRL stars are overplotted as points with arrows representing their proper motions. The color represents the distance offset from the cluster location at $(\phi_1, \phi_2) = (0, 0)\degr$.
    {\it Lower:} Residual density of stars passing an isochrone selection with $m-M = 16.8$ in the region around Pal~13. The histogram on the right illustrates the density summed along $\phi_1$ within $|\phi_1| <  5\degree$, excluding the region within 0\fdg4 of the globular cluster.}
    \label{fig:zoom}
\end{figure*}

\subsection{RR Lyrae Stars}
\label{sec:rrl}

We search for RR Lyrae stars (RRLs) associated with the cluster and tidal tails using the Pan-STARRS1 (PS1) RRL catalog \citep{Sesar:2017c}.
Following the same procedure as in \citet{PriceWhelan:2019}, we perform the search using the 61,795 RRLs selected by \citet{Sesar:2017c} as \textit{bona fide} based on their classification scores, plus the 6,459 \textit{non-bona fide} PS1 stars also identified as RRLs in the \Gaia Specific Objects Study (SOS) catalog.
We adopt photometric distances for the RRLs based on the PS1 (dust-corrected) $i$-band magnitudes \citep[see][for details]{PriceWhelan:2019}; The resulting distances have a precision of $\sim3$\%. We retrieve astrometric information for these RRLs by cross-matching to the \textit{Gaia} DR2 catalog with a 1\arcsec\ tolerance.

Pal~13 is known to have four type \typeab RRLs (V1--V4) \citep{Rosino:1957}, which were recently confirmed as members based on \textit{Gaia} proper motions \citet{Yepez:2019}
Stars V1, V3, and V4 are present in the PS1 RRL catalog, but V2 is missing from both the PS1 and \Gaia DR2 RRL catalogs (likely due to crowding, as it is within 3\arcsec of the cluster center).

We select RRLs with distances (17--35\kpc), positions ($|\phi_2|<1\degr$), and a prograde proper motions (corrected for the solar reflex motion) ($\mu_{\phi_1} > 0$) consistent with Pal~13.
This selection yields a total of eight type~\typeab RRLs: the three stars in the cluster plus another five in the tails.
Of these five, two are likely contaminants as they have proper motions $(\mu_{\phi_1},\mu_{\phi_2})$ close to zero, consistent with the background distribution.
The other three RRLs have $\mu_{\phi_1}$ similar to the cluster RRLs.
Our search does not reveal any significant overdensity of RRLs near the \northernname.

The six likely members (three in the cluster, three in the tails) are shown as large markers in the upper panel of \figref{zoom}.
The arrows represent the proper motion of each star, and the color shows the distance offset in $\kpc$ from the cluster at $(\phi_1, \phi_2) = (0, 0)\degree$.
The \Gaia~DR2 \texttt{source\_id}'s for the RRLs are \texttt{2814894112367752192}, \texttt{2814893910504601600}, \texttt{2811888666052959232} (cluster) and \texttt{2712246494232725120}, \texttt{2815432636842520960}, \texttt{2705495870795198464} (tails).
Using the three RRLs in the cluster, we compute a mean distance of $23.6 \pm 0.2$~kpc.

We use the number of RRLs $N_{RR}$ found to provide a rough estimation of the cluster's initial total luminosity, following the procedure described in \citet{Mateu:2018}, which uses the $\log N_{RR}-M_V$ relation observed for globular clusters and dwarf galaxies. The observed number of 7(3) type~\typeab RRLs in total (tails) yields an estimate $L_V=5.1^{+9.7}_{-3.4}\times 10^3 L_\odot$ ($1.4^{+2.6}_{-0.9}\times 10^3 L_\odot$), slightly higher than the cluster luminosity estimated by 
\citet[$L_V=1.1^{+0.5}_{-0.3} \times 10^3 L_\odot$]{Bradford:2011},  but consistent within the uncertainties. Our findings for the total luminosity  therefore supports the claim made by \citet{Siegel:2001} and \citet{Bradford:2011} that the abnormally large blue straggler population observed in the cluster suggests a higher initial mass and, therefore, luminosity.

\subsection{Stream Model}
\label{sec:model}

The cluster Pal 13 is thought to be on a highly eccentric orbit
\citep[e.g.,][]{Siegel:2001,Vasiliev:2019}, but satellites that disrupt on very
radial orbits are expected to form more diffuse tidal debris structures rather
than linear streams \citep[e.g.,][]{Helmi:1999}.
Is it then surprising that the tidal tails of Pal 13 appear stream-like?
We use the sky track of the tidal tails determined here and the kinematics of
the cluster to fit for the Galactic orbit of the stream, then use the best-fit
orbit to generate simulated tidal debris.

We compile measurements of the Pal~13 mean sky position, proper motion, and line-of-sight velocity---$(\ra, \dec) = (346\fdg685, 12\fdg772)$ \citep{Vasiliev:2019}, $(\mu_{\alpha^*}, \mu_\delta) = (1.615, 0.142) \masyr$ \citep{Vasiliev:2019}, and $v_{\rm los} = 25.9\kms$ \citep{Baumgardt:2019}---and their associated uncertainties.
We adopt a distance to Pal~13 based on the 3 cluster RRLs with PS1 measurements (Section~\ref{sec:rrl}), $d=23.6 \pm 0.2\kpc$.
We adopt recently-compiled Galactocentric solar position and velocity measurements \citep{Drimmel:2018} to transform between heliocentric and Galactocentric quantities.
We use a standard, three-component mass model (implemented in \code{gala}; \citealt{Gala}) to represent the Milky Way, with all disk and bulge parameters fixed to fiducial values \citep{Gala, Bovy:2015}, but we fit for the mass and scale radius of the NFW halo component.
We treat the cluster distance, proper motion, and radial velocity as free parameters and use \texttt{BFGS} optimization \citep{scipy:2001} to maximize the likelihood of the orbit given the data described above.

Our best-fit orbit for Pal~13 has a pericenter of $\sim 9~\kpc$ and an apocenter of $\roughly 69\kpc$ and suggests that the cluster passed through pericenter $\approx 75 \Myr$ ago.
We use this best-fit orbit to run an approximate $N$-body simulation of the stream formation (we do not resolve the internal dynamics or disruption process, but include a mass model for the progenitor cluster and generate tidally-stripped stars following \citealt{Fardal:2015}).
\figref{zoom} (top) shows the sky positions (in stream-aligned coordinates) of star particles from this simulation, colored by relative distance to the cluster.
The candidate RRL stars associated with the Pal~13 tails---identified from proper motions alone---are generally consistent with the distance trend and on-sky distribution of star particles in the stream model.
At most earlier simulation snapshots (i.e. at different orbital phases), the tidal debris tends to be much more diffuse, so the coherence of the tidal tails may be another indicator that the cluster is near pericenter.
This also highlights the importance of present-day orbital phase in determining whether tidal tails are observable.

\section{Discussion \& Conclusions}
\label{sec:discussion}

We have applied an isochrone matched-filter technique to stars in DECaLS to detect evidence for tidal tails coincident with the Pal~13 globular cluster.
This observation complements previous studies, which have shown evidence for tidal disruption of Pal~13, including the recent discovery of an extended low-density halo beyond the Jacobi radius \citep{Piatti:2020}.

The detected tidal tails extend $\roughly 5\degr$ in either direction from the cluster and are well-aligned with the proper motion of Pal~13 \citep{Vasiliev:2019}.
We identify a color--magnitude signal in the on-stream region that is consistent with the stellar population of the Pal~13 cluster.
In addition, we find three new RRL stars likely associated with the tidal tails, along with four RRLs known to be associated with the cluster. 
We generate a model of the tails using an orbit fit to the cluster kinematics and find that the model is consistent with the orientation and spatial distribution of the observed tidal tails, and the distance gradient of the RRLs.

The detection of the Pal~13 tidal tails is complicated by nearby structures in the interstellar dust maps, highlighting the importance of deep, precise photometry and the need for additional phase-space information to confirm and model stellar streams.
Future observations, such as radial velocity and metallicity measurements by \SSSSS \citep{Li:2019}, and deeper photometric and proper motion measurements by the Rubin Observatory \citep{Ivezic:2019}, will allow for higher-precision characterization and modeling of the tidal features of Pal~13.

If confirmed, Pal 13 will be one of only a handful of thin, extended stellar streams that has been confidently associated with a bound progenitor, and, conversely, one of the few globular clusters with detectable long tidal tails.
Future observations by \Gaia, the Rubin Observatory, and WFIRST will allow for the discovery of many more such systems in the coming years.
Increasing the population of streams with known progenitors will provide crucial insight into the tidal disruption of globular clusters, the formation of the stellar halo, and the gravitational field of our Galaxy.

\section{Acknowledgments}

This paper is based upon work that is supported by the Visiting Scholars Award
Program of the Universities Research Association. NS thanks the LSSTC Data
Science Fellowship Program, her time as a Fellow has benefited this work. CM acknowledges support from the DGAPA/UNAM PAPIIT program grant IG100319.

This work has made use of data from the European Space Agency (ESA) mission
{\it Gaia} (\url{https://www.cosmos.esa.int/gaia}), processed by the {\it Gaia}
Data Processing and Analysis Consortium (DPAC,
\url{https://www.cosmos.esa.int/web/gaia/dpac/consortium}). Funding for the DPAC
has been provided by national institutions, in particular the institutions
participating in the {\it Gaia} Multilateral Agreement.

\software{
    astropy \citep{astropy:2018},
    gala \citep{Gala},
    matplotlib \citep{mpl},
    numpy \citep{numpy},
    scipy \citep{scipy:2001}
}

\bibliographystyle{aasjournal}
\bibliography{main}

\end{document}